\documentclass[useAMS,usenatbib]{mn2e}
\usepackage{graphicx}
\usepackage{amssymb}
\usepackage{color}
\usepackage[pdftex,
            colorlinks = true,
            linkcolor = blue,
            urlcolor  = blue,
            citecolor = blue,
            anchorcolor = blue]{hyperref}

\title[Galaxy Evolution in A85]{Galaxy Evolution in the Cluster Abell 85: New Insights from the Dwarf Population}
\author[R. Habas, et al. ]{Rebecca Habas$^{1}$\thanks{E-mail:
rebecca.habas@gmail.com}, Dario Fadda$^{2}$, Francine R. Marleau$^{1}$, Andrea Biviano$^{3}$, \newauthor and Florence Durret$^{4}$\\
$^1$University of Innsbruck, Technickerstrasse 25/8, Innsbruck, A-6020, Austria \\
$^2$SOFIA Science Center, USRA, NASA Ames Research Center, M.S. N232-12, Moffett Field, CA 94035 \\
$^{3}$INAF-Osservatorio Astronomico di Trieste, via Tiepolo 11, 34143 Trieste, Italy \\
$^{4}$UPMC-CNRS, UMR7095, Institut d'Astrophysique de Paris, 98brs Bd Arago, F-75014, Paris, France} 
\begin{document}

\date{Accepted XXXX. Received XXXX.}

\pagerange{\pageref{firstpage}--\pageref{lastpage}} \pubyear{2017}

\maketitle

\label{firstpage}

\begin{abstract}

\noindent We present the first results of a new spectroscopic
survey of the cluster Abell 85 targeting 1466 candidate cluster members within the central
$\sim$1~deg$^2$ of the cluster and having magnitudes $m_r < 20.5$ using VIMOS/VLT and HYDRA/WIYN. A total of 520 galaxies are confirmed as either relaxed cluster members or part of an infalling population. A significant fraction are low mass; the median stellar mass of the sample is $10^{9.6}\,
M_{\sun} $, and 25\%  have stellar masses below $10^9\, M_{\sun}$ (i.e.\, 133 dwarf galaxies). We also identify seven active galactic nuclei (AGN), four of which reside in dwarf host galaxies. 
We probe the evolution of star formation rates, based on H$\alpha$\ emission and continuum modeling, as a function of both mass and environment. We find that more star forming galaxies are observed at larger clustercentric distances, while infalling galaxies show evidence for recently enhanced star forming activity.  Main sequence galaxies, defined by their continuum star formation rates, show different evolutionary behavior based on their mass. At the low mass end, the galaxies have had their star formation recently quenched, while more massive galaxies show no significant change. The timescales probed here favor fast quenching mechanisms, such as ram-pressure stripping. Galaxies within the green valley, defined similarly, do not show evidence of quenching. Instead, the low mass galaxies maintain their levels of star forming activity, while the more massive galaxies have experienced a recent burst.    

\end{abstract}


\section{Introduction}
\label{sec:introduction}

Star formation rates (SFRs) are an important tracer of the
evolutionary history of individual galaxies \citep{Peng10,Schawinski14}. SFRs have been extensively studied and modeled for massive
galaxies, but relatively little is known about the star formation
histories of dwarf galaxies other than in the Local Group
(see {\citealt{Tolstoy09}} for a review). However, understanding the evolution of dwarf galaxies is critical to understanding galaxy formation and evolution in general, as dwarf galaxies are (1) the most numerous galaxies in the universe and (2) the building blocks of more massive galaxies, within the model of hierarchical structure formation.

Although dwarf galaxies in general exhibit higher specific star
formation rates today than more massive galaxies, it is believed that
they are more sensitive to quenching mechanisms that inhibit star
formation. These can take the form of either internal or external
processes.  Star formation, feedback
from supernova or active galactic nuclei (AGN), galactic scale winds,
turbulence, and magnetic fields can, for example, internally regulate
their star formation \citep{Larson74,Schawinski06,Silk98,DiMatteo05}. Many external processes such as ram-pressure
stripping, ultra-violet (UV) radiation, galaxy harassment, or
strangulation can have the same effect \citep{Gunn72,Forbes16,Bialas16,Efstathiou92,Moore96, Larson80}. The relative importance of environment
versus internal regulation for dwarfs is still under debate.

Star formation histories of many dwarfs in the Local Group are well
known and have been modeled in detail using synthetic color-magnitude
diagram modeling for over a decade (a compilation of publications on
the topic is provided by \citealt{Weisz14}). These studies have
revealed a complicated star formation history; dwarf galaxies
typically contain an old stellar population, however the average age
varies tremendously from galaxy to galaxy. Weisz et al.\ find
that 80\% of the stellar content was formed before $z=2$ in dwarfs
with $M < 10^5 M_{\sun}$ but only 30\% of the stars in galaxies with $ 10^7 < M_{stellar}/M_{\sun} < 10^9$ were formed in the same time period. The star
formation often occurs in bursts; there is no definitive explanation for this behavior thus far, but it has been suggested that self-regulation feedback can temporarily halt star formation in dwarfs, which then mimics the oscillating star formation histories observed in some local dwarf galaxies \citep{Stinson07}.

There are also clear environmental trends within the Local Group, however; dwarfs within the
virial radii of the Milky Way or Andromeda have little or no ongoing
star formation and more distant dwarfs are more active \citep{Wetzel15,Slater14,McConnachie12}. The dwarf
spheroidals Cetus and Tucana are two noticeable exceptions to this, but models suggest they have passed within the virial radii
of the Milky Way in the past and their star formation could have been suppressed at that time
\citep{Monelli10}. \citet{Geha12} further demonstrated that
nearly all ($>$99.9\%) isolated dwarfs ($d > 1.5$ Mpc from massive
galaxies) are star forming. This suggests a scenario in which the star
formation in dwarf galaxies is quenched as they move into more dense environments.

In order to achieve better statistics, many studies have extended
observations of dwarf galaxies to nearby clusters. It has been known
for some time that dwarfs obey a morphology-density relation similar
to more massive galaxies \citep{Ferguson90}, but spectroscopic
surveys to probe star formation rates and quenching mechanisms have
only begun recently now that larger telescopes have made these studies
more feasible. In general, these studies find the environment plays a
role in quenching star formation in dwarf galaxies \citep{Darvish16}. For example, \citet{Drinkwater01} found a lack of star forming dwarfs in the
core of the Fornax cluster, while \citet{Mahajan07} observed an increase in star formation in the infalling dwarf galaxy population of the Coma cluster. Star formation rates also appear elevated in dynamically
younger clusters \citep{Lotz03}. In the Virgo cluster, \citet{Cybulski14} found that dwarf galaxies were quenched in smaller groups
before falling into the cluster (`pre-processing') while \citet{Grossi16} demonstrated that gas is stripped from dwarfs as they move
through the cluster environment.

However, there is additional confusion as to whether we are studying
the same systems in clusters as in the field or low density
groups. Dynamically, dwarf galaxies in clusters tend to be less
relaxed than more massive systems \citep{Vijayaraghavan15}; if
this observation is true, however, that begs the question of where the
older dwarf cluster population is. The morphology--density relation
also suggests that there may be a morphological evolution of dwarfs in
clusters; the current favored transition is a late--type spiral that
is stripped of its outer gas and dust until it resembles an
early--type dwarf. Several potential transitions have been put forth
over the years (see, for example, \citealt{Conselice01,Grebel03,Knezek05,Aguerri05}, but there is no conclusive evidence
for any. The point remains, however, that some (or all) of what we
consider a `dwarf' population in cluster environments may be
fundamentally different from dwarfs in the field or Local Group.

More data on star formation in cluster dwarfs is required to understand
the systems that we are studying, and to disentangle the quenching
mechanisms acting on dwarf galaxies. To that end, we performed a new
spectroscopic survey of the $\sim$~1~deg$^2$ region centered on the
cluster Abell 85 (hereafter A85), aimed at studying the low mass
population. A85 is a nearby, moderately rich (R=1) cluster of galaxies
located at $z = 0.055$ \citep{Abell89}. It is a well studied
cluster with over 1000 spectroscopically confirmed cluster members to
date \citep{Agulli14,Ahn12,Bravo-Alfaro09,Durret98a} and references therein). The cluster is
often approximated as a virialized system, with $r_{vir} = 2.8 $~Mpc \citep{Durret05},but many studies show that
A85 is slowly accreting material through several ongoing minor
mergers: a group of galaxies known as the ''Southern Blob" is in an
advanced stage of merging - as evinced by the disruption of the group
- from the south of the cluster \citep{Bravo-Alfaro09}, and
references therein), a second small clump of galaxies is infalling
from the southwest, presumably for the first time \citep{Ichinohe15}, and several kinematically distinct subclumps, at various
stages of disruption, have been identified by \citet{Bravo-Alfaro09} and \citet{Yu16}. In addition, models of the temperature variations in the X-ray emission across the cluster support the picture that A85 has undergone several small mergers/accretion events in the past few Gyr
\citep{Durret05,Ichinohe15}. A85 also has an associated X-ray filament, extending roughly from the Southern Blob to the southeast, in the projected direction of Abell 87 (A87) \citep{Durret98b,Durret05}. It should be noted, however, that A87 is a background cluster at a redshift $z= 0.133$ \citep{Bravo-Alfaro09}, and thus is not physically associated with the filament, which is believed to be oriented perpendicular to our line of sight.

Although the cluster is well studied, star formation rates in the
dwarf galaxy population of A85 remain largely unstudied. One goal of
this work is to measure H$\alpha$\ for a substantial dwarf population
so that we can examine star formation rates across the cluster for a
wide range of masses. This provides a current ($\sim$10~Myr) star formation
estimate, which we compare against the longer star formation timescale
($\sim$100~Myr) given by continuum modeling, which is driven by emission in
the UV \citep{Bell01}. The rest of this paper is organized as follows: Section~\ref{section:data} presents our target selection, datasets, and
observations, Section~\ref{section:results} details the analysis and
results of the cluster membership, mass estimates, star formation
rates, and spatial distribution, Section~\ref{section:discussion}
discusses the results, and finally Section~\ref{section:summary}
summarizes our findings.

To compare our results to previous works, we have adopted throughout
the paper the cosmology: $H_{0}=70$~km~s$^{-1}$~Mpc$^{-1}$, $\Omega_M =
0.3$, $\Omega_{\Lambda} = 0.7$.

\section{Data Sample}
\label{section:data}
The results presented in this paper are based on two spectroscopic
data sets, one taken with Hydra on the WIYN telescope (PI: Fadda,
2008), and the second with VIMOS on the VLT (PI: Marleau,
2013,2014). The target selection criteria and reduction procedures will be
described in greater detail a subsequent data paper (Habas et al.\ 2017, in prep). However, the
main steps for each program are summarized below.

\subsection{WIYN Data}
\subsubsection{Target Selection}
\begin{figure}
\centering
\includegraphics[scale=0.43]{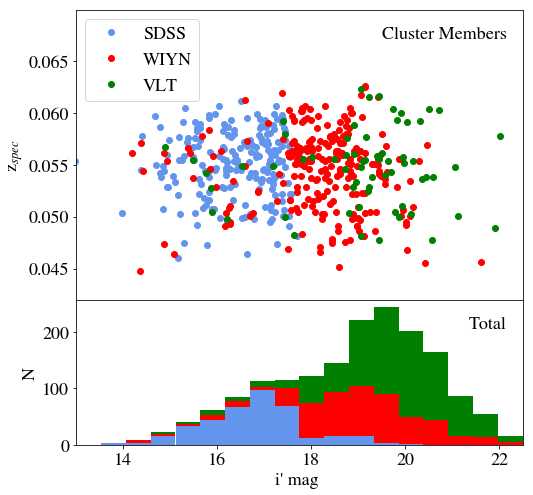}
\caption{{\textit{Top:}} Magnitudes versus redshifts of the A85 cluster members targeted
  by SDSS (blue), WIYN (red; this work), and VLT (green; this work). The cluster membership
  determination is discussed in Section 3.1. {\textit{Bottom:}} distribution of total targets
  as function of their i' magnitudes.}
\label{figure:magnitudes}
\end{figure}

The list of potential targets was compiled from the literature and
SDSS data. The candidates met one of three conditions: (1) the galaxy
was part of observations by \citet{Boue08} and has emission in
a narrow-band H$\alpha$\ filter but an ambiguous redshift determination,
(2) galaxies with a redshift reported in the literature in the range
$0.045<z<0.065$ but no published H$\alpha$\ measurement or star
formation rate, or (3) an SDSS photometric redshift in the range 0.04
$< z <$ 0.1 and coordinates between $0^{\rmn{h}}~40^{\rmn{m}}<\rmn{RA}
< 0^{\rmn{h}}~44^{\rmn{m}}$ and $ -10.4\degr < \rmn{Dec.} <
-8.8\degr$. These criteria generated 91, 142, and 315 targets,
respectively. No magnitude cut was applied to the sample. As can be
seen in Figure~\ref{figure:magnitudes}, the WIYN targets extend the
depth of the SDSS spectroscopically confirmed cluster members by a
few orders of magnitude. Extending this even further was the goal of
our second data set, described in Section 2.2.

\subsubsection{Observations}
The observations were taken with the HYDRA multi-object spectrograph
on the WIYN 3.5m telescope at Kitt Peak National Observatory (KPNO)
during December 2008 and January 2009. We targeted 527 galaxies using
the blue 3.1'' diameter fibers in conjunction with the GG-375 filter
and 600@10.1 grating. In this configuration, the spectra span the
wavelength range 4200 -- 7050 {\AA} and have a dispersion of 1.40
{\AA}/pixel. This filter set ensures coverage of the H$\alpha\ $ and
H$\beta$\ lines for A85 cluster members, so we can estimate star
formation rates corrected for extinction.

\begin{figure}
\centering
\includegraphics[scale=0.38]{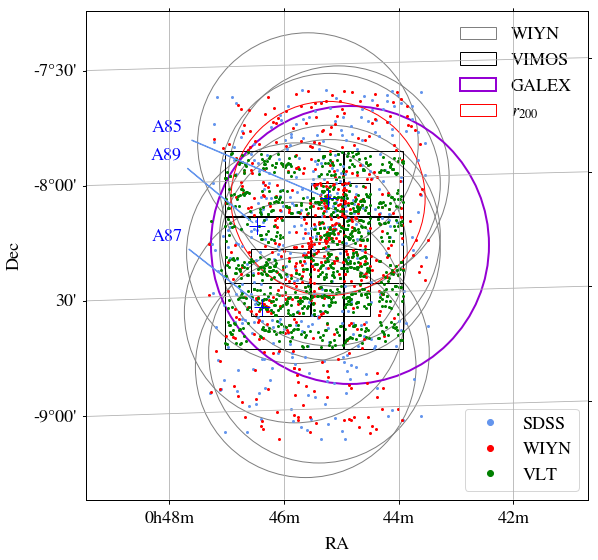}
\caption{ Observing scheme for the two surveys. {Galaxies with SDSS $z_{spec}$ measurements, or targeted by WIYN and VLT are shown as blue, red, and green points, respectively.} The position of
  Abell~85, as well as the positions of the background group(s) and
  cluster of Abell~89 and Abell~87 are marked with
  blue crosses and labelled. The HYDRA/WIYN fibers were placed inside
  the gray circles, while the 12 VIMOS/VLT pointings are shown with
  the black rectangles. We chose to overlap coverage of the
  filament with VIMOS/VLT to better sample the potential cluster members in
  that region. The $r_{200}$, discussed in Section~\ref{section:membership}, is
  shown in red and our GALEX observation in purple.}
 \label{figure:FOV}
\end{figure}

The number of targets required 9 pointings, the outline of which is
shown in Figure~\ref{figure:FOV}, and are summarized in Table~\ref{table:VSPECV}. With HYDRA's 1 degree diameter field of view ($\sim 3.6$ Mpc at the distance of A85), we covered the cluster beyond the virial radius ($r_{vir} = 2.8$~Mpc with our adopted cosmology) estimated by \citet{Durret05}. 
The first eight pointings had exposure times of
3$\times$1ks. The last field exposure was reduced to 2$\times$0.5ks
due to time losses with the Hydra fiber positioner.

\begin{table}
\caption{Summary of targeted fields for both observing campaigns. The
  Hydra observations have a circular, $1^{\degr}$ diameter FOV, while
  the VIMOS fields are composed of four 7\arcmin\ $\times$ 8\arcmin\ quadrants
  arranged in a grid, separated by a 2\arcmin\ gap. These are visualized in
  Figure~\ref{figure:FOV}.  }
\label{table:VSPECV}
\centering
\begin{tabular}{ l l l l  }
\hline
Field & $\alpha$\ (J2000.0)& $\delta$ (J2000.0)& Observation  \\ 
& & &Date (UTC)  \\
\hline
HYDRA 1& 00 42 14.873 & -10 04 01.818 & 2008 Dec 31 \\
HYDRA 2& 00 42 06.802 & -10 00 03.459& 2008 Dec 31\\
HYDRA 3& 00 42 25.764&-09 49 09.59& 2008 Dec 31 \\
HYDRA 4& 00 42 24.910 & -09 32 58.940& 2009 Jan 1\\
HYDRA 5& 00 41 47.58 & -09 32 04.725& 2009 Jan 1\\
HYDRA 6& 00 41 46.997& -09 28 10.366& 2009 Jan 1\\
HYDRA 7& 00 41 37.372& -09 12 01.974& 2009 Jan 2\\
HYDRA 8& 00 41 49.141&-09 14 07.186& 2009 Jan 2\\
HYDRA 9& 00 42 11.89& -09 03 03.363& 2009 Jan 2\\
VIMOS 1 & 00 41 00.0 & -09 14 05.0 & 2013 Oct 03    \\ 
VIMOS 2 & 00 42 05.0 & -09 14 05.0 & 2014 Sep 28    \\ 
VIMOS 3 & 00 43 10.0 & -09 14 05.0 & 2014 Sep 29    \\ 
VIMOS 4 & 00 41 00.0 & -09 32 00.0 & 2013 Oct 03    \\ 
VIMOS 5 & 00 42 05.0 & -09 32 00.0 & 2014 Sep 26    \\ 
VIMOS 6 & 00 43 10.0 & -09 32 00.0 & 2014 Sep 28    \\ 
VIMOS 7 & 00 41 00.0 & -09 50 00.0 & 2013 Aug 04    \\ 
VIMOS 8 & 00 42 05.0 & -09 50 00.0 & 2014 Sep 24    \\ 
VIMOS 9 & 00 43 10.0 & -09 50 00.0 & 2014 Sep 25    \\ 
VIMOS 10 & 00 41 36.3 & -09 23 00.0 & 2014 Sep 27    \\
VIMOS 11 & 00 41 36.3 & -09 41 00.0 & 2013 Sep 10    \\ 
VIMOS 12 & 00 42 41.3 & -09 41 00.0 & 2014 Aug 22   \\ 
\hline
\end{tabular}

\end{table}

\subsubsection{Data Reduction}

The data was reduced using the package {\sl dohydra} by \citet{Valdes95},
the LA-Cosmic package \citep{vanDokkum01} for cosmic-ray rejection, and
custom IDL scripts. We flux calibrated the data using
spectro-photometric standards taken while observing. Since A85 was
low on the horizon during the observations, this calibration efficiently
corrected the shape of the continuum but not the absolute value of the
flux. For this reason, we adjusted the calibration using the SDSS photometry
of the observed targets. This was done using the
\textit{fiberMag\_$<$band$>$} parameter, which contains the magnitude
inside a 3\arcsec diameter fiber centered on each target. As the HYDRA
spectrograph uses nearly identical fibers, no additional adjustments
based on relative sizes or shapes were required.
Calibration factors were obtained for each Hydra configuration using
a median value from all the observed targets.

Redshifts and line properties were extracted using a Python code
written by the authors. An earlier version written in IDL has already
been use to extract parameters from WIYN data (see, for example,
\citealt{Marleau07} and \citealt{Bianconi15}). The updated code is linked
to the SDSS cross-correlation templates to provide more robust,
automated redshift estimates. We were able to extract redshifts for
434 galaxies in the WIYN sample. The code also fits a Gaussian profile
to identified lines, and then extracts line fluxes.

These flux measurements were subsequently corrected for Galactic
extinction, dereddened, and scaled to represent total fluxes for the
galaxies. We adopted a fixed dereddening value $E(B-V)=0.032$ from
\citet{Schlafly11}. The reddening maps from \citet{Schlegel98} do not show much variation across the cluster ($<0.012$ mag
in our targeted region), thus we adopted the extinction measured in
the direction of Holm 15A, the brightest cluster galaxy in A85, for
the entire field. The fluxes were then dereddened for internal
extinction using the standard \citet{Calzetti00} procedure.

Finally, we applied an aperture correction to estimate the H$\alpha\ $
content in the entire galaxy following the prescription outlined in
\citet{Nakamura04}. Motivated by the work of \citet{Ryder94}, they plotted the integrated H$\alpha\ $ and i' band flux
profiles against one another to create an H$\alpha\ $ growth curve for a
sample of galaxies, which appears to be independent of morphologies
(see their Figure~6). Although this growth curve was not designed for
dwarf galaxies, {the shape is expected to be similar for dwarf galaxies;}
\citet{Gallagher89} showed that H$\alpha\ $ in dwarfs tends
to be centrally concentrated, and has a radial profile similar to more
massive galaxies. We can use their composite growth curve to
extrapolate the total H$\alpha\ $ content in our targets given the
measured, central H$\alpha\ $ flux and the ratio of the central i' band
flux (SDSS \textit{fiberMag\_i}) to the total i'-band flux (SDSS
\textit{petroMag\_i}). We have deeper CFHT i'-band imaging (see Section~\ref{section:photometry}), but we did not incorporate those
magnitudes here, as the Nakamura curve was originally normalized using
SDSS data and we did not want to introduce a possible bias by using an
alternative dataset. This procedure generates more physical values
than simply rescaling the aperture to the size of the entire
galaxy. Each galaxy was individually corrected, and the median value
for all objects was 3.95. We identified three outliers (aperture
correction $> 20$); two of which are massive galaxies that are large
enough to require such a large correction. The third object comes from
the fact that our list of WIYN targets included a galaxy that SDSS had
split into two  sources. Therefore, we have removed the additional
source, located at the edge of the galaxy, from our sample.

\subsection{VIMOS Data}

\subsubsection{Target Selection}
With the larger collecting area of the Very Large Telescope (VLT), we focused
on obtaining spectra for potential cluster members at the faint end of
the cluster luminosity function using the VIsible Multi-Object
Spectrograph mounted on UT3 (Melipal). Using $m_r$ as a proxy for
mass, we searched for galaxies in SDSS with $18.0 < m_r < 20.5$ and
$z_{phot}<0.1$. At the redshift of A85, this magnitude range
corresponds to -19.0 $< M_r <$ -16.5 and stellar masses roughly in the
range $ 10^{7.5} M_{\sun }< M_{stellar} < 10^{10} M_{\sun}$. Various cuts
have been used in the literature to define a dwarf galaxy sample (see
\citealt{Dunn10} for a review); if we define our dwarf galaxies with a
stellar mass cut of $10^{9} M_{\sun}$, we anticipate a large fraction
of A85 cluster members will fall in this category. The upper magnitude
limit was imposed in order to limit our targets to a number that could
reasonably be observed in each VIMOS FOV. Near the cluster outskirts,
where the target density was lower, we also included objects of
interest detected in the far-IR Herschel bands as fillers.

\subsubsection{Observations}
We targeted 918 galaxies that met the above criteria with VIMOS
 on the 8.2 meter VLT. The data was taken
using the MR GG475 grism and 1\arcsec\ width slits. Compared with WIYN,
the VLT has better spatial resolution, but the MR grism has slightly
lower spectral resolution. The spectra have a dispersion of 2.5
\AA/pixel, and cover the wavelength range 4800 -- 10000~\AA.

VIMOS has a smaller field of view (16\arcmin\ $\times$ 18\arcmin\ with a
2\arcmin\ gap between quadrants), and we required more pointings to
achieve similar coverage to the WIYN data.  Rather than covering the
entire cluster, we chose to overlap pointings in the southern half of
the cluster, to ensure good sampling in the filament region and areas
of ongoing mergers. The VIMOS mapping can be seen in Figure
~\ref{figure:FOV}, and the coordinates and dates are summarized in
Table~\ref{table:VSPECV}. Although we do not span as wide an area with VIMOS as the WIYN observations, the VIMOS coverage extends from the cluster center to the south beyond $r_{200}$, the radius inside which the mean density is 200 times the critical density of the universe and which is expected to enclose most of the virialized cluster mass. Due to bad weather, only four fields were observed in fall 2013, while the remaining fields were observed in fall 2014. Two quadrants were unusable due to poor atmospheric conditions.

\subsubsection{Data Reduction}
The VIMOS data was reduced using the ESO recommended Reflex
software. We made minor adjustedments to the {\it dispersion}, {\it reference
wavelength}, {\it start wavelength}, {\it wdegree}, and {\it wreject} parameters to
better reflect our setup, but all other parameters were left with
their default values. Line fluxes and redshifts were extracted using
the same code that was used for the WIYN data after adapting it for
the Reflex outputs. We were able to extract redshifts for 780 galaxies
of the 875 usable spectra taken with the VLT, giving a $\sim$90\% success rate.

Aside from the flux calibration, the extracted line fluxes were
corrected following a similar procedure as the WIYN data, described
above. This procedure was more complicated than for the WIYN/HYDRA
data, as we first had to convert the flux in the slit to an equivalent
fiber flux. To accomplish this we extracted the flux inside the
(1\arcsec\ $\times$ 2.46\arcsec) rectangular extraction region used by
Reflex, centered on our target coordinates, and compared this with the
flux extracted from our r' band pre-imaging data using a 2\arcsec\
diameter circular region centered on the object. Each galaxy was
individually corrected by the corresponding multiplicative factor. The
median value for all objects was 1.32 with a standard deviation
0.16. This corrected magnitude was compared to the SDSS
\textit{fiber2Mag\_r} parameter, the magnitude within a 2\arcsec\
diameter fiber centered on the object, to compute the flux calibration
factor. We computed a median value, 1.48, which we applied to all of
the targets. Overall, these two corrections resulted in a median
correction of 1.95.

The subsequent dereddening and aperture corrections follow the same
procedure as the WIYN data, described above, but utilizing the
\textit{fiber2Mag\_$<$band$>$} parameters from SDSS where required.

\subsection{Photometric Data}
\label{section:photometry}
This paper also makes use of several photometric data sets in order to
flux calibrate the spectra and estimate stellar masses. In the optical
we utilized SDSS (DR7 for the target selection, described above:
\citealt{Abazajian09}; DR12, for flux calibration and estimating the
stellar masses, see Section~\ref{MAGPHYS}: {\citealt{Alam15}})
u',g',r',i',z' photometry and deep, archival u,g,r,i band images taken
with the Canada-France-Hawaii Telescope (CFHT; \citealt{Boue08}). The CFHT images have limiting magnitudes of 27.3, 27.5,
26.4, and 26.0, respectively, and a resolution of 0.187\arcsec\/pixel, making
them significantly deeper than the SDSS imaging which reaches limiting
magnitudes $\le 22.2$ in each band. However, the CFHT imaging only
covers the southern region of the cluster, and we required the SDSS
data for complete optical photometric coverage of our spectroscopic
targets. Along with the optical imaging, we included MIR data from
WISE, specifically the SDSS/WISE forced photometry \citep{Lang14}, as well as our own GALEX NUV and IR WIRC Palomar data. The photometric
properties and data reduction of our GALEX and WIRC-Palomar
observations will be described in detail in Habas et al.\ (2017, in
prep).

\section{Results}
\label{section:results}
\subsection{Cluster Membership}
\label{section:membership}

\begin{figure*}
\centering
\includegraphics[scale=0.42]{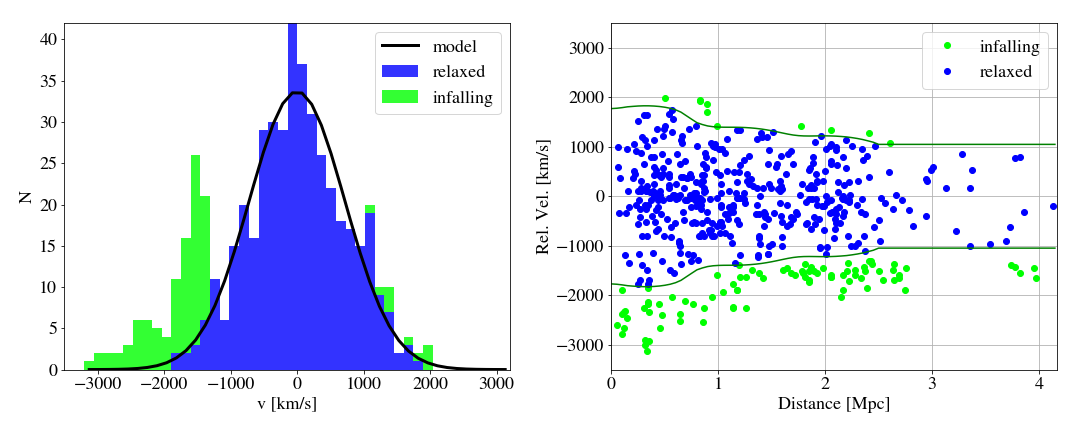}
\caption{{\textit{Left:}} The distribution of cluster members in velocity space, as returned by the {\textit{shifting gapper}} technique. The relaxed and infalling populations were identified based on the positions of the galaxies in relative velocity vs. clustercentric distance space and a model of the cluster. The cluster redshift and one of the velocity dispersions reported in Section~\ref{section:membership} are based on the Gaussian used to model the cluster population (solid line). {\textit{Right:} The positions of the galaxies in relative velocity vs. clustercentric distance space. The solid lines mark the separation between the relaxed cluster population and the infalling galaxies, and is from the modeled mass profile of the cluster \citep{Gifford13a,Gifford13b}.} }
\label{figure:veldist}
\end{figure*}

\begin{figure}
\centering
\includegraphics[scale=0.36]{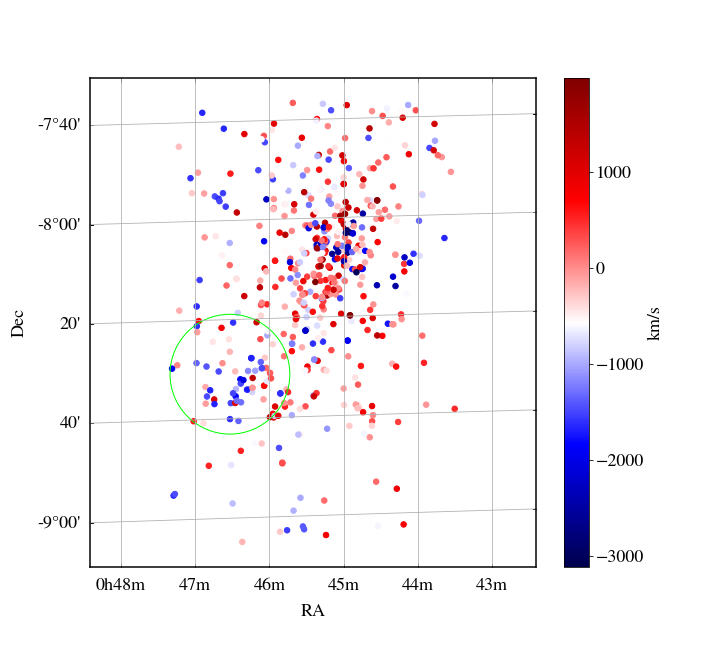}
\caption{Two-dimensional spatial distribution of A85 cluster members
  and infalling galaxies. The points are color coded by their
  velocity, relative to the cluster center. We tentatively identify
  one foreground group, indicated by the green circle.
}
\label{figure:spatial}
\end{figure}

{The cluster membership was determined using 
the {\it shifting gapper} technique from \citet{Fadda96}.} This method removes near field interlopers
based on the galaxies' velocities and clustercentric distances. We
used the default bin size of 500~kpc from the cluster center (or wide
enough to contain 15 galaxies), but increased the default gapsize from 1000~km/s to 1150~km/s. {Other groups have reported velocity dispersion greater than 1000~km/s for A85 (1122~km/s; \citealt{Bravo-Alfaro09}) due to the unrelaxed state of the cluster, and we chose a slightly higher value to ensure we did not miss any members.} A histogram of the velocity dispersion of 
the 520 galaxies returned by this technique is shown in Figure~\ref{figure:veldist}, {\textit{left}}. There is a distinct foreground population of galaxies at $v_{cluster} -
v_{group} \sim 1600 $~km/s, which is similar to the relative velocities of
infalling groups found in other clusters \citep{Cortese06}. Spatially, however, this peak does not belong to a single, well defined group. Rather, these galaxies are spread across the targeted area, with a slight overdensity to the south-east of the cluster center, along the filament. This region is highlighted by the green circle in Figure~\ref{figure:spatial} and corresponds well with the position of the SE group identified in \citet{Bravo-Alfaro09}. These galaxies with higher relative velocities will be referred to as infalling galaxies for the remainder of the paper.

{To isolate the relaxed cluster members from the infalling population,} we subsequently applied the caustic
algorithm\footnote{https://github.com/giffordw/CausticMass} presented
in \citet{Gifford13a} and \citet{Gifford13b} to the
cluster members identified above. This method independently checks for and removes likely cluster interlopers, models the mass profile of the galaxy, and defines the edge of the cluster through the average escape velocity, defined from an iso-density surface in $r-v$ space. The interlopers (in this case, the infalling galaxies) are rejected by a second, more aggressive application of the {\it shifting gapper} technique to remove the foreground peaks in the velocity dispersion in  Figure~\ref{figure:veldist}. The mass of the cluster is then calculated following the methodology of \citet{Diaferio99}, which can model the mass profile of clusters to large radii where 
 the system is no longer in dynamical equilibrium. The caustic identified 433 relaxed cluster members. Based on this subsample, the modeled cluster has $r_{200} = 1.73^{+0.03}_{-0.04}$ Mpc, mass enclosed within $r_{200}$ is estimated to be $M_{200} = 6.1^{+0.3}_{-0.4} \times 10^{14} M_{\sun}$, and a cluster velocity dispersion of $\sigma = 782^{+4}_{-21}$~km/s. The errors were estimated based on jackknife resampling of the dataset. The velocity dispersion is lower than values often reported in the literature \citep{Girardi97,Bravo-Alfaro09,Sifon15}, and is due to the separation of the
infalling galaxies; applying a simple Gaussian profile to the velocity
distribution of the main peak in Figure~\ref{figure:veldist} returns
a similar value (762.3~km/s) to the caustic returned dispersion, while the inclusion of the infalling galaxies
gives a value closer to that in the literature (1004~km/s). 
Our values for $r_{200}\ $ and $M_{200}\ $ similarly show some tension with published measurements, which again depends on the treatment of the infalling galaxies and subgroups within the cluster. In a study of the dynamical state and substructures of several clusters, \citet{Wen13} estimate a similar $r_{200}\ $ (1.68~Mpc) for A85, while other groups have calculated values closer to 2~Mpc \citep{Durret05,Sifon15}. Regarding the mass, Sifon et al.\ calculated a value $M_{200} = 10 \times 10^{14} M_{\sun}\ $ ($r_{200} = 2.3$~Mpc) for A85; Girardi et al.\ find a similar value within $r=1.5$~Mpc, but only when they fail to reject substructures. After isolating the main cluster population, they found a revised mass similar to ours, $M_{<1.5~\mbox{Mpc}} = 6.5 \times 10^{14} M_{\sun}$ for A85. From this subsample of relaxed cluster members, we further estimate $z=0.0557\pm0.0001$, which is in good agreement with other values in the literature (see Figure 1 of \citealt{Yu16} for a review of previous measurements).

{For the remainder of this paper, we discuss the population of all 520
cluster members identified via the initial {\it shifting gapper} method. We use the results of the
caustic modeling to identify galaxies within three subpopulations:} the core (members
inside the caustic and $r < r_{200}$; 300 galaxies), cluster outskirts (members
inside the caustic and $r > r_{200}$; 121 galaxies), and the infalling population
(the galaxies outside the caustic boundary; 99 galaxies).

\subsection{Stellar Mass Estimates}
\label{MAGPHYS}
We estimated the stellar masses of the cluster members using the
MAGPHYS (Multi-wavelength Analysis of Galaxy Physical Properties) code
by \citet{Cunha08}. We used NUV to MIR photometry described above
corrected for Galactic extinction in each bandpass using the values
from \citet{Schlafly11} to obtain the spectral energy
distribution (SED). MAGPHYS models the emission of stars and the
interstellar medium using 50,000 stellar population spectra and 50,000
dust emission spectra, joined in a physically consistent manner to
simulate the galaxies' SEDs, and returns the physical parameters from
the best model. We used the default cosmology in the fits, which
corresponds to the one adopted in this paper (see
Section~\ref{sec:introduction}). 

We obtained stellar masses in units of solar masses ranging from 6.88 to 11.29 in log space, with a median value of 9.58. A `traditional' dwarf cut at $\log (M_{stellar}/M_{\sun}) < 9.0$ yields 133 galaxies, or 25\% of our sample. {We did not reject model parameters for any of our galaxies; we visually inspected the MAGPHYS output, and the model fits and  $\chi^2$ values were reasonable. }

\subsection{AGN in cluster members}

\begin{figure}
\centering
\includegraphics[scale=0.4]{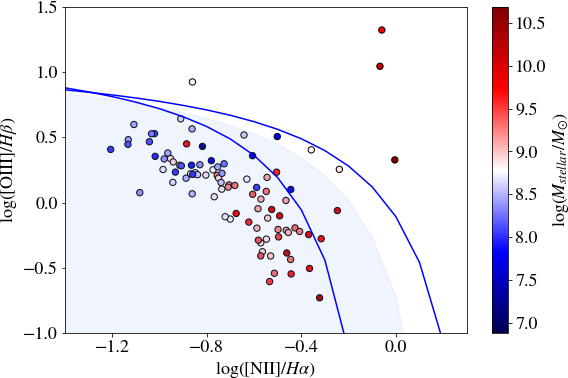}
\caption{A BPT diagram showing the 97 A85 cluster members for which we
  could measure all required line fluxes. The solid lines mark the
  boundaries between star forming, composite, and AGN dominated
  regions as defined by \citet{Kewley01} (upper line) and \citet{Kauffmann03b} (lower line); we were less restrictive, however, and excluded the
  three galaxies in the composite region nearest the AGN demarcation
  in addition to the four AGN candidates. All galaxies in the blue
  shaded region were retained in our analysis of star formation
  rates. The stellar mass of each source is shown with a colour gradient.}
\label{figure:bpt}
\end{figure}

Before calculating the star formation rates (see Section~\ref{section:sfr}) of the A85 cluster members, we removed those
galaxies whose emission is dominated by an 
AGN. We separated these from the star forming galaxies based on
their positions in the [OIII](5007~\AA)/H$\beta$\ versus [NII](6584~\AA)/H$\alpha$\ BPT diagram \citep{Baldwin81}. We were able to measure all required lines
for 80 of the WIYN ($\sim$~18\%) and 17 ($\sim$~24\%) of the VLT
targets. The results can be seen in Figure~\ref{figure:bpt}. The
solid lines are from \citet{Kewley01} and \citet{Kauffmann03b}, and designate regions of star formation dominated
emission, a composite region, and AGN dominated emission. Four of our
galaxies fall in the AGN dominated region; however, we also dropped
three galaxies that fall just inside the composite region. All
galaxies within the blue shaded region of Figure~\ref{figure:bpt}
were retained in the following analysis of star formation rates.

Interestingly, half of the AGN in the sample reside in dwarf galaxy
hosts; the three galaxies in the composite region as well as the
left-most AGN dominated galaxy have host stellar masses $< 10^9 M_{\sun}$.

\subsection{Star Formation Rates}
\label{section:sfr}

\begin{figure*}
\centering
\includegraphics[scale=0.42]{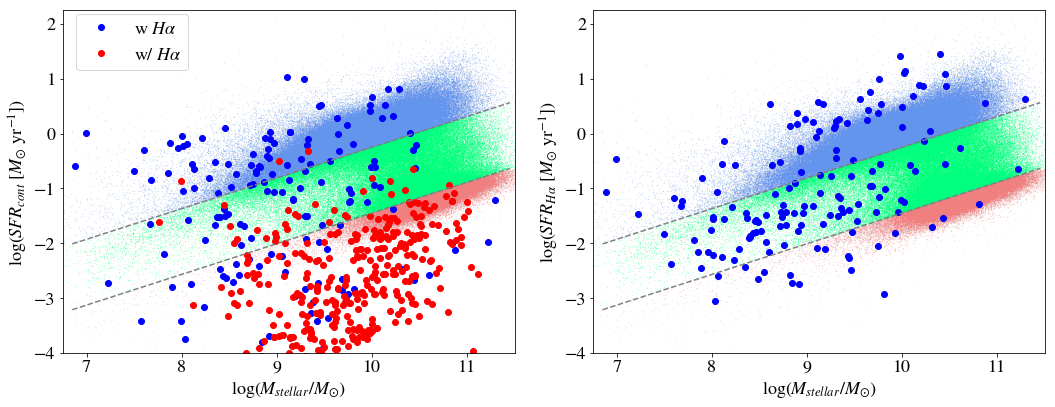}
\caption{The star formation main sequence (SFMS) based on stellar masses returned by MAGPHYS and our two star formation estimators. {\textit {Left:}} SFRs calculated from modeling the  continuum
  emission. Galaxies that coincidentally have H$\alpha$\ measurements
  are shown in blue, while all other galaxies are plotted in red. {\textit{Right:}} SFRs calculated from H$\alpha$\
  emission. {The error of the H$\alpha$ flux measurements is the same size as the points.}  In both
  plots, our data points are plotted over the \citet{Brinchmann04} sample of field galaxies, restricted to objects with $z <
  z_{A85} + 0.03$, and color coded blue, green, and red according to their position on the main sequence, green valley, or red cloud. The separation between the
  main sequence and green valley is assumed to be at 0.3~dex lower than the
  fit to the main sequence and the separation between the green valley and red cloud is 1.5~dex lower; both are denoted by the grey dashed lines.}
\label{figure:Halpha_sfr}
\end{figure*}

{Star formation rates were estimated using the total, corrected H$\alpha$\ fluxes and assuming a Kroupa initial mass function (IMF; \citealt{Kroupa02,Calzetti13}):}
\begin{center}
\begin{equation}
\mbox{SFR}_{H\alpha} \mbox{[M}_{\sun} \mbox{yr}^{-1}\mbox{]} = 5.5 \times 10^{-42} L(H\alpha) ~ [\mbox{ergs s}^{-1}]
\end{equation}
\end{center}
{A Kroupa IMF further assumes that stars are formed with masses between $0.1 - 100
M_{\sun}$, and that the star formation is roughly constant on
timescales of $\tau \geq6$ Myr.} Only 30\% of the cluster members (158 galaxies) show H$\alpha$\
emission. The resulting star formation rates, plotted against stellar
mass, are shown in Figure~\ref{figure:Halpha_sfr}, \textit{right}.

{The median error in the H$\alpha$\ flux measurements is only $\sim$10\%, and therefore the error in the calculated star formation rates is dominated by systematic uncertainties in the corrections we applied; the flux correction adds a factor of $\sim2$ (this is less for the WIYN targets) to the flux, the dereddening procedure adds a median correction factor of $\sim3.3$, and the median aperture correction is $\sim4$. The aperture correction, in addition to being the largest correction, is the most uncertain of the three, particularly for the low mass galaxies. However, the aperture correction therefore has less of an impact on our population of dwarf galaxies than the more massive galaxies; we find that the size of this correction varies with the stellar mass of the galaxy, varying from 2.7 for galaxies with $\log(M_{stellar}/M_\odot) < 8$ to 5.3 for galaxies with $\log(M_{stellar}/M_\odot) > 10$. This is not unexpected, since massive galaxies at a fixed distance will typically have larger angular sizes.  }
 
{For comparison, we extracted star formation rates and stellar mass estimates, taking median values from the model fitting, for a population of field galaxies from the online MPA-JHU
catalogues\footnote{http://wwwmpa.mpa-garching.mpg.de/SDSS/DR7/}, based on SDSS DR7 data.}  The
stellar mass estimates available through the website were obtained by
modeling the broad band SDSS photometry (\textit{modelMag\_$<$band$>$}
parameters) using the stellar population synthesis codes of \citet{Bruzual03}. The methodology is similar to that of \citet{Kauffmann03a} and \citet{Gallazzi05}, although these authors use
fits to spectral features rather than broadband photometry.  The star
formation rates are calculated following a similar technique to that
discussed in \citet{Brinchmann04}; classes of galaxies were
identified (e.g.\ AGN, star forming, or low S/N), and emission lines
were modeled using the \citet{Charlot01} models, from which
star formation rates were extracted. Two updates to the methodology
have been applied: the dust attenuation is now accounted for in the
probability distribution function which removes all trends depending
on the dust, and fit stochastic models to the photometry of the
galaxies to obtain a better aperture correction, following the method
of \citet{Salim07}. These models adopt the same initial mass function (Kroupa) and cosmology assumed in this paper. {This data was used to locate the main sequence, green valley and red cloud at $z_{A85}$. We found a linear best fit to the main sequence (hereafter referred to as BFMS) for the subsample of galaxies with $|z -
z_{A85}| < 0.03$; restricting the galaxies used in the fit reduces the scatter found in the full sample and allows us to avoid any bias due to the evolution of the main sequence with redshift. Once we found the BFMS, we applied vertical offsets of -0.3 dex and -1.5 dex to roughly follow the inner edges of the upper and lower overdensities, respectively, thus defining the three regions. } 

{The demarcation lines, as well as data for individual galaxies in the MPA-JHU catalog with $z < z_{A85}+0.03$ are also plotted in Figure~\ref{figure:Halpha_sfr}. It should be noted that at the redshift of A85, there is not a significant population of field dwarf galaxies in SDSS; although there is some evidence that the main sequence may not be linear in all regimes (e.g. \citealt{Whitaker14, Erfanianfar16}), there is little evidence for non-linear behavior in the local universe (e.g. \citealt{Peng10,Brinchmann04,Speagle14}) and we assume the BFMS extends to galaxies with stellar masses below $10^9M_\odot$. The time dependent star formation main sequence presented by \citet{Speagle14} suggests that there is very little change in the BFMS between $z=0$ and $z=0.055$, with the slope of the main sequence changing by $<0.2$~dex and the zeropoint by $<0.1$~dex. Based on the regions defined by the BFMS, 52 of the 151 cluster members (after removing the AGN) with measured H$\alpha\ $ emission lie on the main sequence, 77 inhabit the green valley, and 22 fall in the red cloud.}

Although we have GALEX NUV imaging of the cluster, we lack FUV
coverage, and therefore cannot estimate a SFR based on FUV emission
directly. By including this data in MAGPHYS, however, the models can
better constrain the star formation on timescales of
$\sim$~100~Myr. We extracted these values to compare against the more
recent star formation indicated by H$\alpha$\ emission (timescales $\sim$~10~Myr). This can be
seen in Figure~\ref{figure:Halpha_sfr}, \textit{left}. The MAGPHYS estimated SFRs place 64 galaxies on the main sequence, 72 in the green valley, and 380 galaxies within (or below) the red cloud. The blue points are the
MAGPHYS estimated star formation rates for galaxies that also have
H$\alpha$, while the red points are those without H$\alpha$\
measurements. With a few exceptions, the two populate the expected
regions of the plot: the galaxies with H$\alpha$\ inhabit the main sequence and green valley, depending on the strength of the H$\alpha$\ emission, while galaxies with H$\alpha$\ non-detections are most heavily concentrated within the red cloud.

\section{Discussion}
\label{section:discussion}

{In order to trace the change in SFRs over time, both as a function of galaxy stellar mass and environment, we focused on the subsample of galaxies with measured H$\alpha$\ emission and continuum SFRs that place them either in the continuum main sequence (61 galaxies) or the continuum green valey (48 galaxies). As mentioned earlier, the continuum SFRs returned by MAGPHYS hinge on the UV flux of the galaxy and probes timescales on the order of 100 Myr, while the H$\alpha$\ emission probes star formation within the past $\sim$10 Myr; comparing these will allow us to examine changes in the galaxy SFRs. The galaxies within the continuum red cloud were disregarded, as too few have H$\alpha$\ emission that could be used for comparison. }

\begin{figure*}
\centering
\includegraphics[scale=0.45]{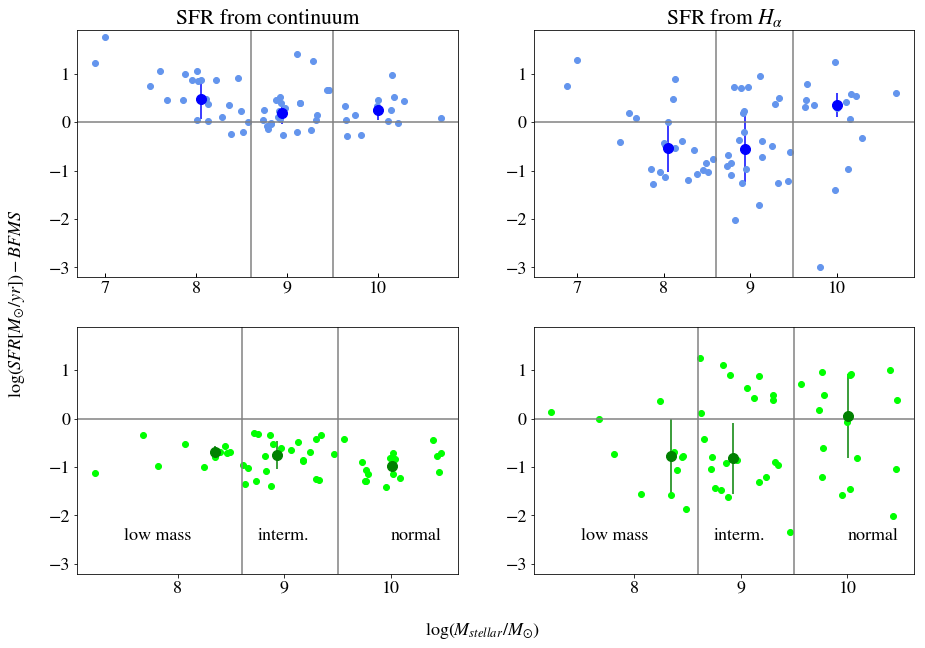}
\caption{The relative star formation rates of individual galaxies compared to the best fit to the main sequence (BFMS). {\textit{Top:}} The subset of 66 galaxies that fall within the main sequence based on their continuum SFRs, and have H$\alpha$\ measurements (see Figure~\ref{figure:Halpha_sfr}). These are split into relative star formation rates based on the continuum SFR ({\textit{left}}) and H$\alpha$\ SFR ({\textit{right}}). {\textit{Bottom:}} The same figures as above, but for the sample of 48 galaxies that fall within the green valley based on their continuum star formation rates. A horizontal line at $\log (SFR) - BFMS$ = 0 is plotted to help guide the eye. Each plot is further subdivided into three mass bins, inside which the galaxies appear to exhibit different behaviors. The large points in each bin represent median values of the mass and star formation rate, with the error bars showing the dispersion in the relative star formation. }

\label{figure:sfr-ms}
\end{figure*}

The populations we retained for this analysis are isolated in Figure~\ref{figure:sfr-ms}. The SFRs are plotted relative to the BFMS line described in Section~3.4, to provide a baseline comparison with typical field values per unit stellar mass. The relative SFRs are plotted for continuum SFRs ({\textit{left}}) and H$\alpha$\ SFRs ({\textit{right}}) and galaxies on the continuum main sequence ({\textit{top}}) and within continuum the green valley ({\textit{bottom}}). In order to explore any dependence on stellar mass, we further divided the galaxies into three mass bins: galaxies with $\log(M_{stellar}/M_{\sun}) < 8.6$, $ 8.6 \leq \log(M_{stellar}/M_{\sun}) < 9.5$, and $\log(M_{stellar}/M_{\sun}) \geq 9.5$, which we will hereafter refer to as `low mass', 'intermediate mass', and `normal mass' galaxies, respectively. The mass cuts were chosen to divide the sample into bins with roughly even statistics and do not correspond to any traditional dwarf/non-dwarf cut. On the main sequence, the low mass, intermediate, and normal mass bins contain 23, 24, and 15 galaxies, respectively, while the corresponding mass bins in the green valley contain 11, 22, and 16 galaxies. To help illustrate trends in the data, median SFRs from each bin are overplotted as larger points in Figure~\ref{figure:sfr-ms}. The error bars show the dispersion in the relative star formation rates, which was calculated using a median absolute dispersion (MAD) estimator. 

{The distribution of SFRs for galaxies on the continuum main sequence and green valley (Figure~\ref{figure:sfr-ms}, left plots) are statistically distinct from the H$\alpha$\ derived SFRs in these galaxies (Figure~\ref{figure:sfr-ms}, right plots). We applied an Anderson-Darling test for normality, which is more sensitive to the tails of the distribution than the standard KS test \citep{Feigelson12}. We can reject the null hypothesis (i.e., that the continuum and H$\alpha$ star forming populations are sampled from the same parent population) at the 5\% significance level for the entire sample of galaxies on the continuum main sequence and the continuum green valley ($A=15.76$ and $A=6.34$ respectively). We also applied the test for the galaxies within each continuum main sequence mass bin; the null hypothesis is rejected at the 5\% level for galaxies within the low mass and intermediate mass bins ($A=11.45$ and $A=7.15$, respectively), \ {making the trends summarized by the median value in the low and intermediate mass bins significant}. We cannot, however, reject the null hypothesis for the normal mass galaxies ($A=0.36$). 

\ {The dispersion in the H$\alpha$ SFRs for individual galaxies is much larger than the continuum population, as expected. H$\alpha$ is sensitive to bursts of star formation on short timescales and how completely the IMF is sampled (which can be an issue for galaxies with low L$_{H\alpha}$), making it a less stable indicator of SFRs. Rather than tracing the star formation history of individual galaxies, we instead focused on any trends superimposed atop the scatter in H$\alpha$ SFRs, as highlighted by the median values in Figure~\ref{figure:sfr-ms}; the continuum main sequence SFRs indicate that the $\sim$100 Myr star formation was comparable or slightly higher than star formation levels in similar field galaxies, but the H$\alpha$\ emission indicates that the SFRs have since fallen, on average, in the low and intermediate mass galaxies. Statistically, no conclusions can be drawn for the normal mass galaxies. }

\begin{figure*}
\centering
\includegraphics[scale=0.42]{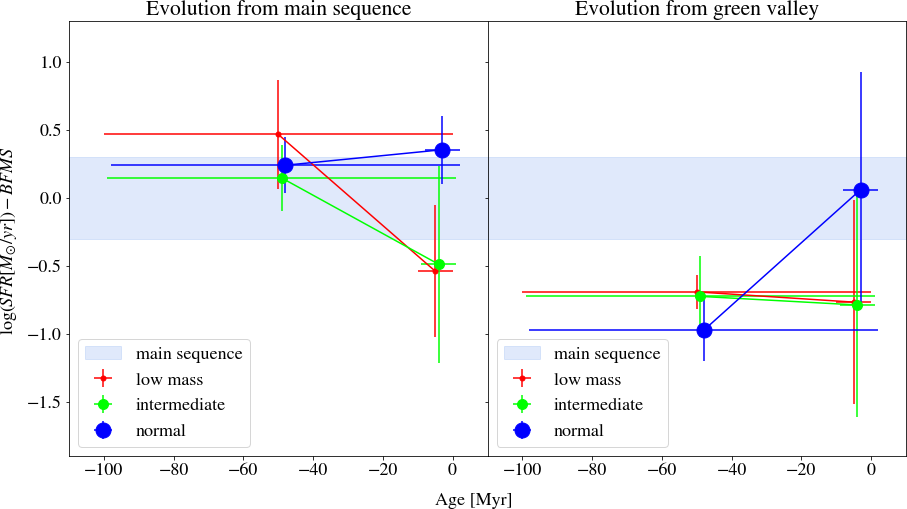}
\caption{ The evolution of galaxies initially on the continuum main sequence (\textit{left}) and continuum green valley (\textit{right}), plotted explicitly as a function of time. The data points are the median $\log$(SFR)$-BFMS$ points taken from Figure~\ref{figure:sfr-ms}, and are used to show the overall trends from each mass bin. We cannot assign an explicit age to the galaxies, so the points are plotted near the median age of each indicator (50~Myr for $SFR_{cont}$ and 5~Myr for $SFR_{H\alpha}$); they are slighly offset from one another so the vertical error bars (the dispersion of the SFR in the respective bin) can be clearly seen. The horizontal error bars show the timescales probed by the two SFRs. The populations on the continuum main sequence and continuum green valley show very different evolutionary trends when compared with the ongoing SFRs in the respective galaxies. On average galaxies on the continuum main sequence have maintained their rates of star formation or been suppressed within the past $\sim$100~Myr, while galaxies originally in the continuum green valley have either maintained their SFRs or experienced a recent boost in star formation. The trends in the low and intermediate mass galaxies are similar, although the dispersion in the SFRs of the intermediate mass galaxies increases at recent times. }
\label{figure:SFR-MS-ev}
\end{figure*}

{The change in star formation rates for the continuum green valley galaxies is noticeably different from the continuum main sequence galaxies described above. Applying the Anderson-Darling test within each mass bin, we cannot reject the null hypothesis that the continuum and H$\alpha$ star forming populations are subsamples of the same parent (normal) population at the 5\% level for the low and intermediate mass galaxies ($A=0.14$ and $A=1.76$), but can for the normal mass galaxies ($A=3.72$). Interestingly, the overall trend in the normal mass bin is not that the median SFR has been further suppressed. Rather, it has been boosted to the level of star forming, main sequence galaxies in the field. The median values in the low and intermediate mass bins are consistent with no evolutionary changes, although once again individual galaxies show a large variations between the continuum and H$\alpha$\ SFRs. The dispersion in these bins increased by a factor of $\sim$6 and $\sim$3, respectively, in the H$\alpha$\ plot. This behavior counters models in which galaxy evolution is a simple transformative process that moves galaxies unidirectionally from the main sequence to the red cloud. As we show in Figure~\ref{figure:veldistSFR} and discuss later in the paper, this enhancement is likely due to galaxy-galaxy interactions or mergers as the galaxies fall into the cluster.  }

{
It has been suggested recently that H$\alpha$\ may not be as reliable of a star formation tracer in dwarf galaxies as FUV emission, which is directly observable \citep{McQuinn15,Lee09,Fumagalli11}. The work by \citet{Lee09} determined that H$\alpha$\ underpredicted the star formation in local (d$<$11~Mpc) dwarfs when compared to estimates returned from FUV luminosities; such a finding would have a direct impact on the results presented above, as this is exactly the trend that we observe for galaxies within the continuum main sequence. \citet{Lee09} present an empirical, recalibrated Kennicutt H$\alpha$\ SFR \citep{Kennicutt98}, that will convert non dust corrected L($H\alpha$) measurements to FUV level star formation rates for objects with galaxy-wide  L($H\alpha$$)<10^{39}$~erg/s (see their equation 10). We recalibrated the equation for a Kroupa IMF and applied it to our data to test the impact of potentially underpredicted H$\alpha$\ SFRs. This only affected 13 galaxies in our continuum MS sample (7 in the low mass bin, 6 in the intermediate mass bin) and 15 galaxies in the continuum green valley sample (7 in the low mass bin, 8 in the intermediate mass bin). The recalibrated SFRs for these galaxies typically differed by $<0.3$ in log space. Because this was calculated based on non dust corrected luminosities, the correction could be positive or negative, and the overall scatter seen in Figure~\ref{figure:sfr-ms} does not vary significantly. In light of this, we still believe the trends observed in this work are real.  
}

The differences between the two timescales and three mass bins are explicitly plotted in Figure~\ref{figure:SFR-MS-ev} for the galaxies on the continuum main sequence (\textit{right}) and those within in the continuum green valley (\textit{left}). Each data point corresponds to one of the median values from each mass bin and each emission source probed in Figure~\ref{figure:sfr-ms}. The relative star formation rates are now plotted as a function of time; we cannot assign definitive ages to the galaxies and simply chose to plot the points at the median time associated with the two indicators ($\sim$50~Myr for SFR$_{cont}$ and $\sim$5~Myr for SFR$_{H\alpha}$), albeit slightly offset from one another so the vertical error bars can be distinguished. The x-axis error bars show the range of times probed by the UV ($\sim$0 -- 100~Myr) and H$\alpha$\ emission ($\sim$0 -- 10~Myr), respectively. The y-axis error bars show the same dispersion as in  Figure~\ref{figure:sfr-ms}. For the population on the continuum main sequence, the normal mass galaxies show, on average, a stable rate of star formation on these timescales with no evidence of evolution. The low and intermediate mass galaxy populations have both experienced quenching during this timeframe, suppressing the median SFR in both of these mass bins to levels below the main sequence. This effect is most pronounced in the low mass galaxies, which initially had a median SFR above the main sequence defined in the Brinchmann et al.\ data (shaded blue in Figure~\ref{figure:SFR-MS-ev}) but has been suppressed at recent times to a value under the main sequence. The difference in timescales probed by the continuum modeling and H$\alpha$\ emission suggests that this evolution is, on average, a rapid process. This may be due to ram-pressure stripping, which simulations suggest can quench star formation within 200 Myr \citep{Steinhauser16}, although the efficiency of this process on dwarf galaxies has been called into question \citep{Emerick16}. Our data does not support quenching mechanisms that act on the order of Gyr, such as strangulation \citep{Peng15} or galaxy harassment \citep{Sensui99}, although we cannot rule out a scenario in which a fast process initially quenches the star formation to low levels, which is ultimately terminated via a slow quenching process. \ {Any environmental effects are superimposed on stochastic star formation histories of individual galaxies, which complicates the above interpretation (e.g., \citealt{Fumagalli11,Kelson14}).}

\begin{figure}
\centering
\includegraphics[scale=0.33]{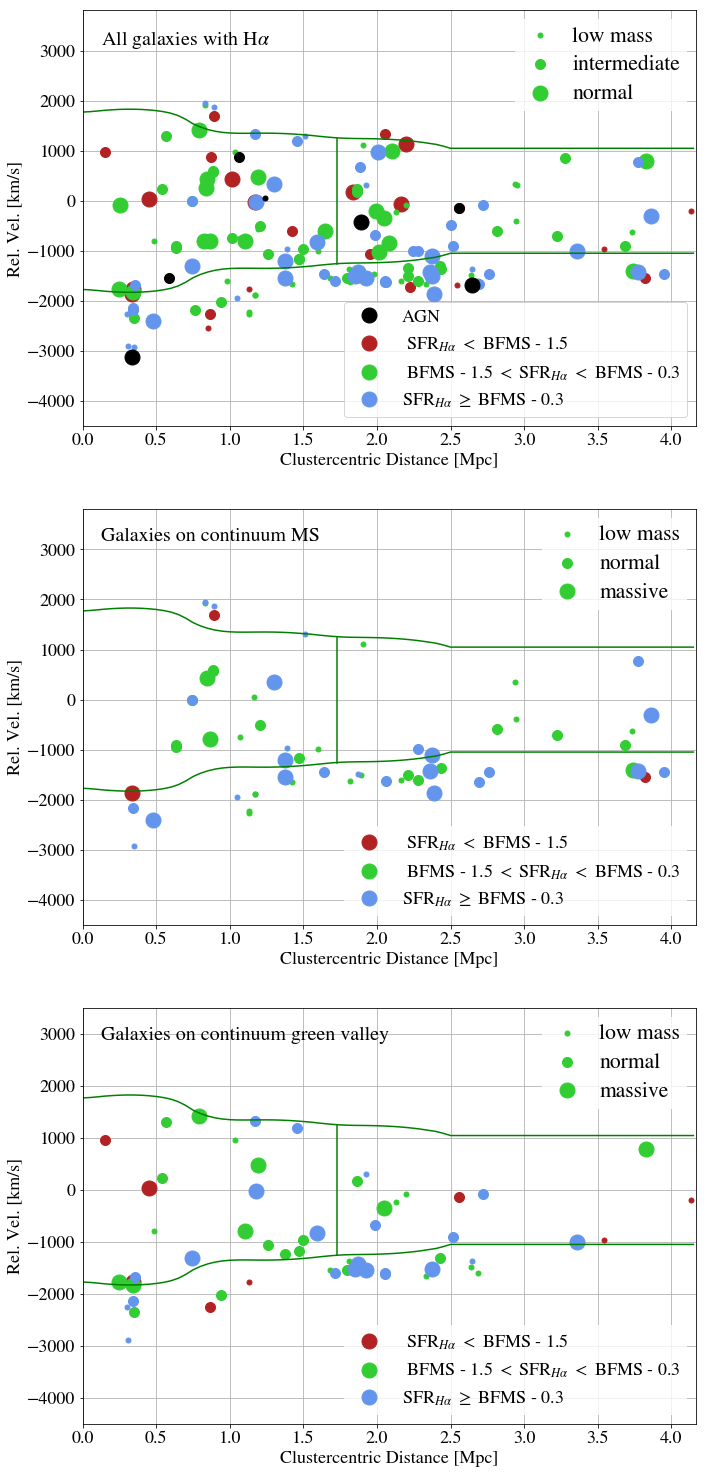}
\caption{\textit{Top:} The distribution of the 158 galaxies with H$\alpha$\ in relative velocity - clustercentric distance space. {The
  galaxies are color coded according to their position within $\log(SFR)$~vs.~$\log(M_{stellar})$ space}, while the AGN candidates are plotted in
  black. The points are further plotted in three sizes, corresponding to the 
  `low mass', `intermediate' , and `normal' mass bins defined earlier. The caustic edges and $r_{200} = 1.73$~Mpc are also displayed, to separate the core, outskirts, and infalling regions. Passive galaxies (those without H$\alpha$\ detection) are omitted in the plot, but are also present throughout all regions of the cluster (see Figure~\ref{figure:figureadded}. \textit{Middle:} The distribution of the 61 galaxies originally on the continuum main sequence. The colors are the same as above; the blue galaxies have maintained their SFRs, while the green points have seen been suppressed. \textit{Bottom:} Same as the above, but for the distribution of 48 galaxies within the continuum green valley. The blue points have experienced a recent increase in star formation}
\label{figure:veldistSFR}
\end{figure}

\begin{figure*}
\centering
\includegraphics[scale=0.35]{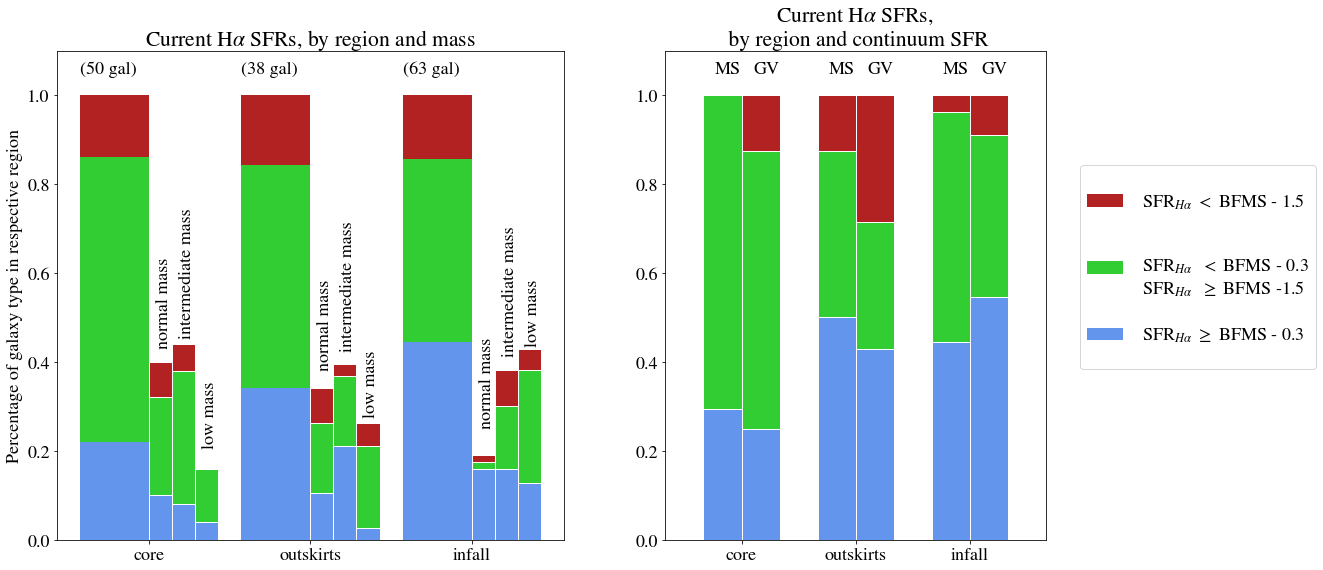}
\caption{\textit{Left:} The distribution of galaxies with H$\alpha$\ emission by environment and mass. The wider bins display the star formation rates for all galaxies within that environment, normalized to one to compare the relative fraction of galaxies on the main sequence (blue), in the green valley (green), or red cloud (red). The narrower bins show the SFR breakdown within each environment according to the stellar mass of the galaxies. \textit{Right:} The evolution in star formation rates according to environment. Each environment has two bars representing galaxies originally on the continuum main sequence (labeled MS) or continuum green valley (labeled GV), which are color coded according to the levels of ongoing SF as measured from H$\alpha$\ emission. Each bar is again normalized to one, to compare relative fractions. No more than 50\% of the continuum MS galaxies have continued to form stars at such high rates in any environment. The continuum GV galaxies that have experienced a recent boost the H$\alpha$\ MS can be found in all environments, but are predominantly located among the cluster outskirts and the infalling population. }
\label{figure:figureadded}
\end{figure*}

In contrast, the galaxies in the continuum green valley show strikingly different behavior. In no mass bin do the galaxies show a significant decline in recent SFRs; rather, the low and intermediate mass galaxies have, on average, maintained their levels of star formation and the normal mass galaxies have experienced a recent boost. Other groups have observed similar enhancements in galaxy SFRs in clusters \citep{Poggianti04,Porter07}, and generally attribute this to galaxy-galaxy harassment as they enter the more crowded infall region. \citet{Mahajan12} also found this enhancement is more prominent in dynamically unrelaxed clusters (such as A85), as the local overdensities of the infalling substructures and the low velocity dispersion contribute to the number of galaxy-galaxy interactions.

To explore whether or not this enhancement occurs in infalling galaxies, we also examined the spatial distribution of the galaxies with ongoing star formation, seen in Figure~\ref{figure:veldistSFR}. {The distribution of the AGN and all 158 galaxies with H$\alpha$\ emission are shown in the top plot, while the middle and lower plots isolate the galaxies originally on the continuum MS and GV, respectively. The data points are colored according to their current(H$\alpha$) SFRs and scaled according to mass, to visualize the combined effects of environment and stellar mass on the galaxies' star formation histories. Here we also explicitly include the galaxies from the continuum main sequence and green valley that have little H$\alpha$\ emission, placing them within the red cloud currently, which were ignored previously. The cluster boundaries, identified by the caustic, are marked in green, as is $r_{200}$.  The relative number of galaxies, broken down according to environment, mass, and continuum star formation rates, are directly compared in Figure~\ref{figure:figureadded}.}

{Galaxies that were in the continuum green valley but which have since experienced a boost in star formation can be found in every environment probed here. However, they represent an increasing fraction of the galaxies as one moves from the cluster core to the cluster outskirts and then the infalling population. The number of galaxies in each region is too low to argue that this enhancement is statistically significant, but it is suggestive.}

{We also examined the spatial distribution of the AGN candidates. They are preferentially found SE of the cluster, along the filament, although this is a projection effect as no clustering is observed in $r-v$ space. They tend to be found at larger distances from the cluster center, with 5/7 of the AGN at projected distances $1 - 2.65$~Mpc. this is in good agreement with other studies that find higher AGN fractions at distances $\sim 1 - 2 r_{200}$ \citep{Bianconi15, Haines13, Pimbblet13}. Interestingly, all four of the AGN located in dwarf galaxy hosts are among the relaxed cluster members, and the three least massive hosts ($\log(M_{stellar}/M_{\odot}) < 8.85$) are located within the cluster core.   }

It should not be forgotten, however, that passive galaxies - those with no H$\alpha$\ emission - dominate our sample, and are present in all three mass bins (14 low mass galaxies, 124 intermediate mass galaxies, and 224 normal mass passive galaxies) in all three environments (247 in the core, 81 in the outskirts, and 34 of the infalling galaxies). {These galaxies are not shown in Figures~\ref{figure:veldistSFR} and \ref{figure:figureadded}, however, as they overwhelm the star forming galaxies we wanted to highlight.} In each of the three environments probed, the fraction of passive galaxies increases with increasing mass. Similarly, within each mass bin, the fraction of passive galaxies increases as one moves to more dense environments. In our sample, only 10\% of the infalling dwarfs are passive, while 88\% of the normal mass galaxies in the core have no measured H$\alpha$\ flux. This mimics trends observed in other clusters (e.g. \citealt{Haines06, Bianconi15}). Pre-processing of some form is required to explain the presence of the passive galaxies in the outermost regions. Studying three clusters of galaxies, \citet{McIntosh04} found that galaxy morphological transformations have already begun in accreted galaxies before other processes such as galaxy harassment or ram-pressure stripping come into play, and therefore must occur at large distances from the cluster center. Similarly, \citet{Gomez03} found suppressed star formation in galaxies at distances as large as $3 - 4 r_{vir}$, when compared with a field population. Galaxies may also be quenched in smaller groups before reaching the cluster, but we can place no constraints on these mechanisms with the present study.

\section{Summary and Conclusions}
\label{section:summary}
This project had two primary objectives: to push the
magnitude limit of known A85 cluster members to fainter values and thereby
identify a substantial dwarf galaxy cluster population, and to obtain spectra
with the resolution and wavelength range sufficient to estimate star
formation rates of cluster members from the Hydrogen Balmer lines. 
To that end, we targeted 1466 galaxies within roughly 1 square degree field centered on
Abell 85 in a spectroscopic survey using both WIYN/HYDRA and the
VLT/VIMOS.

\begin{itemize}
\item We identified {520 cluster members in A85. After isolating the relaxed cluster members (433 galaxies)}, we measured $z = 0.0557 \pm 0.0001$ and obtained more robust estimates of the velocity dispersion and $r_{200}$ for the cluster than previously reported: $782^{+4}_{-21}$~km/s and $1.73^{+0.03}_{-0.04}$~Mpc, respectively.

\item Stellar masses were estimated with MAGPHYS using fluxes from NUV to MIR. The sample covers almost 4.5 orders of magnitude in mass in log space; 133 of the 520 galaxies included in this study are identified as dwarf galaxies, using a stellar mass cut of $M_{stellar} = 10^{9} M_{\sun}$.

\item We tested the source of the measured emission on a BPT diagram for the 20\% of our sample with sufficient line measurements. From this, we identified seven
  galaxies with a significant AGN contribution that we dropped from
  the subsequent star formation analysis. Four of these reside in dwarf galaxy
  hosts. The AGN are preferentially located at distances of $1 - 2.5$~Mpc from the cluster center, in good agreement with other studies.

\item SFRs were estimated two ways: (1) from H$\alpha$\ fluxes, corrected for extinction and scaled to represent the total  H$\alpha$\ flux in each galaxy, and (2) extracted from
  MAGPHYS modeling of the continuum. These estimate the SFRs within the past $\sim$10~Myr and $\sim$100~Myr, respectively, allowing us to probe \ {recent changes in the star formation histories of our targets. Individual galaxies show a range of behaviors; the SFRs in some galaxies remain nearly unchanged, some have experienced a recent boost, while others have been quenched within this timescale.  }

\item \ {The galaxies were further subdivided into three stellar mass bins (low, intermediate, and normal), and we explored changes in the median star formation rates within each mass bin. We interpret these overall trends as an evolutionary effect.} The galaxies in the low and intermediate mass bins have evolved similarly, although the observed dispersion in the intermediate mass galaxies is larger. Those that were on the continuum main sequence have SFRs that have been, on average, suppressed in recent times; the timescales probed by our SFR estimators favor a fast quenching process, such as ram-pressure stripping, over slower mechanisms that act on the order of Gyr. The low and intermediate mass populations in the continuum green valley show no evidence for evolution in the past $\sim$100~Myr. In contrast, the galaxies within the normal mass bins show markedly different behaviors; we find no difference is SFRs for the galaxies that were on the continuum main sequence, while those in the continuum green valley show recent enhancements in their star formation rates.

\item Galaxies with measurable H$\alpha$\ emission reside in all environments that we probed -- core, cluster outskirts, and among the infalling galaxies -- although the highly star forming galaxies are preferentially located in less dense environments, as expected. {We further explored the evolutionary trends of the galaxies on the continuum main sequence and within the continuum green valley in conjunction with their spatial distribution; the data suggests that the galaxies that have moved from the continuum green valley to the H$\alpha$\ main sequence are more common in the cluster outskirts and the infalling population, although our statistics are too low to make a firm conclusion. If true, however,} this would support a scenario in which more frequent galaxy-galaxy interactions have caused a recent, temporary burst of star formation. It should also be noted that passive galaxies with no H$\alpha$\ emission are also located in all environments. It is possible that these galaxies were pre-processed in groups before moving closer to the cluster, or that the cluster environment affects infalling galaxies at distances much larger than $r_{200}$. 
\end{itemize}

\ {
How applicable these results are to other clusters remains to be seen. The environment of Abell 85 appears fairly typical of a relatively relaxed cluster; previous works have identified several substructures within Abell 85, and the cluster appears to have been gradually built through a series of minor mergers (at least over the past few Gyr). However, the number of substructures in other clusters, relative masses of the cluster and its substructures, as well as the infall history and timescales may all affect the results found in this paper.  }

\section*{Acknowledgments}
We thank Tom Jarrett \& Louise Edwards for their help in reducing the
near-IR images. We would also like to thank the anonymous referee for their help improving the paper. This work is based in part on observations made with
GALEX, a space telescope operated by the Jet Propulsion Laboratory,
California Institute of Technology under a contract with NASA. Support
for this work was provided in part by NASA through an award issued by
JPL/Caltech.

Funding for SDSS-III has been provided by the Alfred P. Sloan
Foundation, the Participating Institutions, the National Science
Foundation, and the U.S. Department of Energy Office of Science. The
SDSS-III web site is http://www.sdss3.org/.

SDSS-III is managed by the Astrophysical Research Consortium for the
Participating Institutions of the SDSS-III Collaboration including the
University of Arizona, the Brazilian Participation Group, Brookhaven
National Laboratory, Carnegie Mellon University, University of
Florida, the French Participation Group, the German Participation
Group, Harvard University, the Instituto de Astrofisica de Canarias,
the Michigan State/Notre Dame/JINA Participation Group, Johns Hopkins
University, Lawrence Berkeley National Laboratory, Max Planck
Institute for Astrophysics, Max Planck Institute for Extraterrestrial
Physics, New Mexico State University, New York University, Ohio State
University, Pennsylvania State University, University of Portsmouth,
Princeton University, the Spanish Participation Group, University of
Tokyo, University of Utah, Vanderbilt University, University of
Virginia, University of Washington, and Yale University.

\label{lastpage}

\end{document}